\documentclass{ws-procs9x6}

\def\bsbsb{$B_s^0$-$\overline{B}{}_s^0$}
\def\cA{{\cal A}}
\def\Ga{\Gamma}
\def\bb{\overline{B}{}}
\def\dabs{(\Delta a^{B_s}){}}

\newcommand{\refeq}[1]{(\ref{#1})}
\def\etal {{\it et al.}}

\begin{document}

\title{D\O\ EVIDENCE FOR CP VIOLATION
AND IMPLICATION FOR CPT VIOLATION IN $B$-MESON MIXING}

\author{R.\ VAN KOOTEN$^*$ FOR THE D\O\ COLLABORATION$^{**}$}

\address{Department of Physics, Indiana University\\
Bloomington, Indiana 47405, USA\\
$^*$E-mail: rvankoot@indiana.edu\\
$^{**}$http://www-d0.fnal.gov }

\begin{abstract}
A D\O\ analysis measuring the charge
asymmetry $A^b_{\mathrm{sl}}$  
of like-sign
dimuon events due to semileptonic $b$-hadron decays
at the Fermilab
Tevatron Collider is described.  
It differs by 3.2 standard deviations from the
Standard Model prediction to provide first 
evidence of CPT-invariant anomalous CP violation in the mixing
of neutral $B$ mesons, and is compared to the CP-violating
phase obtained from a D\O\ analysis of the time-dependent decay
angles in $B^0_s \rightarrow J/\psi \phi$.  If CPT violation
is allowed, the dimuon asymmetry also yields the first 
sensitivity to CPT violation in the $B^0_s$ system. 
\end{abstract}

\bodymatter

\section{Introduction}

The interferometric systems of
the particle-antiparticle oscillations of neutral mesons are
particularly sensitive to testing for CP and CPT violation.
CP violation has been observed at small levels in a number of
these
systems\cite{review_CP}, and the violation of CP symmetry is
a necessary condition for the matter-antimatter asymmetry of 
our universe\cite{baryogenesis} and for our very existence.
However, the observed levels 
of CP violation in the $K^0$  and
$B^0_d$ systems are not large enough to account for this asymmetry,
implying the need for additional sources of CP violation beyond the
Standard Model (SM).
Both CP and CPT violation in the neutral $B$-meson system are
considered below.


\section{D\O\ dimuon charge asymmetry}

In 6.1~fb$^{-1}$ of $p\bar{p}$ collision data, the
D\O\ Collaboration first measures\cite{D0_pubs} the raw
dimuon charge asymmetry
$A = (N^{\mu^+\mu^+} - N^{\mu^-\mu^-})/(N^{\mu^+\mu^+} + N^{\mu^-\mu^-})$ 
regardless of muon source. From pure physics processes at the
primary interaction, one of the very few sources of same-sign 
dileptons in the same collision event is due to $B$ physics. 
If there is a nonzero asymmetry after correcting for backgrounds,
the assumption is that it is coming from neutral $B$-meson mixing, i.e., 
the dimuon charge asymmetry of semileptonic $B$ decays
$
A^b_{\mathrm{sl}} = (N_b^{\mu^+\mu^+} - N_b^{\mu^-\mu^-})/(N_b^{\mu^+\mu^+} + N_b^{\mu^-\mu^-}).
$
The great majority of $b$ quark production at the Tevatron
is via $b\bar{b}$, so this can occur, for example, when the $b$ quark
decays semileptonically directly
$\overline{B}{}^0_q \rightarrow \mu^-$, but for the $\bar{b}$ quark, there
is first a $B$-meson oscillation before the semileptonic
decay, i.e., $B^0_q \rightarrow \overline{B}{}^0_q \rightarrow \mu^-$. 
Another way to measure this asymmetry is via inclusive `wrong-sign' decays, 
i.e., $\overline{B} \rightarrow \mu^+ X$ which is only possible  through
flavor oscillation of $B^0_d$ and $B^0_s$. A semileptonic charge
asymmetry can then be constructed:
\begin{equation}
a^b_{\mathrm{sl}} =
\frac{\Gamma(\overline{B} \rightarrow \mu^+) - \Gamma(B \rightarrow \mu^- X)}
{\Gamma(\overline{B} \rightarrow \mu^+) + \Gamma(B \rightarrow \mu^- X)},
\label{eq:absl}
\end{equation}
and probed by measuring the inclusive raw single muon asymmetry
$a = (n^{\mu^+} - n^{\mu^-})/(n^{\mu^+} + n^{\mu^-})$.
Assuming CPT symmetry holds, it can be shown\cite{equala} that
$A^b_{\mathrm{sl}} = a^b_{\mathrm{sl}}$.
Both the raw asymmetries contain contributions from $A^b_{\mathrm{sl}}$,
other processes producing muons, plus detector related backgrounds.
The background contributions to the asymmetries are mostly 
determined by independent measurements in the data, with 
minimal input from simulation. 
Monte Carlo simulation is only used to determine the
remaining fraction of like-sign dimuons,
and fraction of single muons 
from mixed $b$-hadron decays.
The raw asymmetry $a$ is dominated by backgrounds, 
and the different signal and correlated background content of
both asymmetries is used to minimize the total uncertainty on 
$A^b_{\mathrm{sl}}$.

From $1.5 \times 10^9$ muons in the inclusive sample,
$a = (0.955 \pm 0.003)\%$, and from $3.7 \times 10^6$ events
in the like-sign dimuon sample, $A = (0.564 \pm 0.053)\%$.
The most important detector-related background is having one muon
from a semileptonic $b$-hadron decay, and the other from a 
decay in flight of $K \rightarrow \mu \nu$ and $\pi \rightarrow \mu \nu$,
subsequent punch-through to the muon detectors (from showers in 
the material of the calorimeters), and sail-through of $\pi$, 
$K$ and $p$, where
these hadrons are either from the other $b$-hadron decay or from
fragmentation. 
The polarities of the D\O\ detector toroid and solenoid magnets
are switched every two weeks during data running, so that residual
muon reconstruction charge asymmetries cancel to first order.
This reduces these background asymmetries from ${\cal{O}}(3\%)$ 
to less than $0.1\%$. Without this capability, the systematic uncertainty
on the final asymmetry would be substantially larger.

From the like-sign dimuon sample, after correcting for backgrounds,
\begin{equation}
A^b_{\mathrm{sl}} = -0.00736 \pm 0.00266 \thinspace 
{\mathrm{(stat)}}\pm 0.00305 \thinspace {\mathrm{(syst)}}.
\label{eq:dimuona}
\end{equation}
The corrected asymmetry from the inclusive single muon sample is
used to constrain the backgrounds in the dimuon sample and a linear
combination chosen to minimize the total uncertainty on $A^b_{\mathrm{sl}}$
arriving at:
\begin{equation}
A^b_{\mathrm{sl}} = -0.00957 \pm 0.00251 \thinspace 
{\mathrm{(stat)}}\pm 0.00146 \thinspace {\mathrm{(syst)}}.
\label{eq:dimuonf}
\end{equation}
This result is 3.2 standard deviations away from the SM
prediction for CPT-preserving T violation, which is\cite{SMpred}
$A^b_{\mathrm{sl}}{\mathrm{(SM)}} = (-2.3^{+0.5}_{-0.6}) \times 10^{-4}$ and 
represents the first evidence for anomalous CP violation in the mixing
of neutral $B$ mesons.

\section{Comparison to CP-violating phase in $B^0_s$ system}

\begin{figure}
\psfig{file=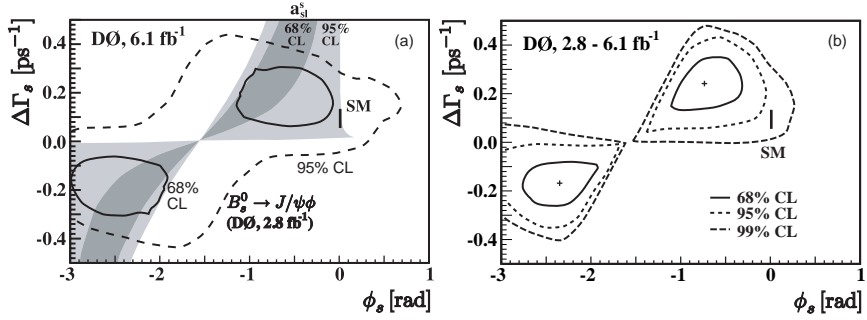,width=4.5in}
\caption{(a) The 68\% and 95\% C.L. regions (shaded bands) 
of probability for $\Delta\Gamma_s$ and $\phi_s$ from
the dimuon asymmetry compared to the regions (solid and dashed
lines) from the $B^0_s \rightarrow J/\psi \phi$ measurement.
(b) Combination of the two results and resulting C.L. regions.
The SM prediction for these parameters is also shown for both
cases.
}
\label{fig:phis}
\end{figure}

The asymmetry above has contributions from both $B^0_d$ and 
$B^0_s$ oscillations, and using world-average values of
mixing parameters\cite{review_CP,D0_pubs}, 
\begin{equation}
A^b_{\mathrm{sl}} =
(0.506 \pm 0.043)a^d_{\mathrm{sl}} +
(0.494 \mp 0.043)a^s_{\mathrm{sl}}.
\label{eq:combo}
\end{equation}
If the world-average\cite{review_CP} 
value of $a^d_{\mathrm{sl}}$ from the
$B$ factories running at the $\Upsilon(4S)$ is input to the
above,  the value $a^s_{\mathrm{sl}} = (-1.46 \pm 0.75)\%$
is obtained, less than three standard deviations from the 
SM prediction\cite{SMpred} 
of $a^s_{\mathrm{sl}}(SM) = (-0.0021 \pm 0.0006)\%$
due to the uncertainty on the coefficients of
Eq.\ \refeq{eq:combo}. 
However, this still provides an interesting comparison
to a D\O\ result\cite{d0jpsiphi} where CP violation is probed 
by the time-dependent angular analysis of decay products
in $B^0_s \rightarrow J/\psi \phi$ where there is interference
between the diagrams with and without ($\overline{B}{}^0_s \rightarrow
J/\psi \phi$) mixing. 
The connection is through the expression\cite{SMpred}
$a^s_{\mathrm{sl}} = (\Delta\Gamma_s/\Delta M_s)\tan\phi_s$ 
where $\Delta\Gamma_s$ and
$\Delta M_s$ are the width and mass differences between mass
eigenstates for the $B^0_s$, and 
$\phi_s$ is the measured CP-violating phase angle between 
the complex off-diagonal mixing-matrix elements
$\Gamma_s^{12}$ and $M_s^{12}$.
The comparison shown in Fig.~\ref{fig:phis} shows
an intriguing similar trend in difference from the SM prediction
of the two analyses.

\section{Implications for CPT violation}

A CPT-violating effect in $B$-meson mixing
was predicted some time ago
\cite{kp}
as potentially arising from spontaneous breaking of Lorentz symmetry
in an underlying unified theory
\cite{ksp},
and the \bsbsb\ system is of particular interest
for studies of CPT violation
because several complete particle-antiparticle oscillations
occur within a meson lifetime
\cite{d0cdf}.
Working
within
the comprehensive effective field theory
describing general Lorentz violation at attainable energies
known as the Standard-Model Extension (SME)
\cite{ck},
each CPT-violating term in the SME Lagrange density
is the product of a CPT-violating operator
and a controlling coefficient, and in this case a combination
of the four SME coefficients\cite{ak3} $(\Delta a^{B_s})_\mu$.

For these purposes, the asymmetry using 
single inclusive muons of Ref.\ \refcite{D0_pubs} and
the combined asymmetry of Eq.\ \refeq{eq:combo} are irrelevant, 
and only the first asymmetry 
$A^b_{\mathrm{sl}}$
result of Eq.\ \refeq{eq:dimuona}
is considered.
A measure of CPT violation is given by the 
inclusive `right-charge' muon charge asymmetry
$\cA^b_{\rm CPT}$
of semileptonic decays of $b$ hadrons,
\begin{equation}
\cA^b_{\rm CPT} =
\frac{ \Ga(\bb \to \mu^- X) - \Ga(B \to \mu^+ X) }
{ \Ga(\bb \to \mu^- X) + \Ga(B \to \mu^+ X) }.
\label{asymmCPT}
\end{equation}
In terms of this CPT asymmetry and the T asymmetry
of Eq.\ \refeq{eq:absl},
\begin{eqnarray}
A^b_{\rm sl} &=& 
{
\displaystyle
\left( \frac{1 + a^b_{\rm sl}}{1 - a^b_{\rm sl}}
-
\frac{1 + \cA^b_{\rm CPT}}{1 - \cA^b_{\rm CPT}}
\right) \Big/
\displaystyle
\left(
\frac{1 + a^b_{\rm sl}}{1 - a^b_{\rm sl}}
+
\frac{1 + \cA^b_{\rm CPT}}{1 - \cA^b_{\rm CPT}}
\right)
}
\approx
a^b_{\rm sl} - \cA^b_{\rm CPT} . 
\quad
\end{eqnarray}
Assuming the only source of T violation
is the SM contribution
$a^b_{\rm sl} {\rm (SM)} = A^b_{\rm sl} {\rm (SM)}$,
combining with the D\O\ dimuon asymmetry of
Eq.\ \refeq{eq:dimuona}
results in
\begin{equation}
\cA^b_{\rm CPT} =
0.00713 \pm 0.00405.
\label{asymmexpt}
\end{equation}
To interpret this as a measure of CPT violation in
$B$-meson mixing, and in particular in the \bsbsb\ system,
the $\omega \xi$ formalism\cite{ak3} is adopted that
allows for CPT violation of arbitrary size, governed
by a complex parameter $\xi$ of arbitrary size.
The complex parameter $\xi$ cannot be a scalar since
CPT violation comes with Lorentz violation\cite{owg}.
It must depend on the $B$-meson four-momentum and is 
therefore a frame-dependent quantity.
The rotation of the Earth relative to the constant
vector $\Delta \vec a$ generates a variation
with sidereal time in $\xi$, 
but it is possible to average over the sidereal time 
and the meson four-momentum. 
Since the particle distributions from $b$-hadron
decay for the Tevatron collider are symmetric in
local detector polar coordinates for D\O, the
dependence 
on the spatial components $(\Delta a^{B_s})_J$ cancels.
Details of the relationship between 
the asymmetry $\cA^b_{\rm CPT}$, 
the parameter $\xi_s$ specific for the \bsbsb\ system, and
the time component SME coefficient $(\Delta a^{B_s})_T$
can be found in Ref.\ \refcite{cpt_preprint}.
Assuming that the only source of CPT violation
comes from \bsbsb\ mixing, the value
\begin{equation}
\dabs_T =
(3.7 \pm 3.8) \times 10^{-12} {\rm ~GeV},
\label{result}
\end{equation}
is found. 
This corresponds to the bound
\begin{equation}
- 3.8 \times 10^{-12} < \dabs_T < 1.1 \times 10^{-11}
\label{confint}
\end{equation}
at the 95\% confidence level.
The value of Eq.\ \refeq{result}, documented in Ref.\ \refcite{cpt_preprint},
represents the first sensitivity to CPT violation
in the \bsbsb\ system.

\section*{Acknowledgments}
The D\O\ Collaboration thanks the staffs at Fermilab and collaborating
institutions, and acknowledge support from agencies including the DOE
and NSF (USA).


\begin{thebibliography}{99}

\bibitem{review_CP}
C.\ Amsler \etal, Phys. Lett. B {\bf 667}, 1 (2008), and
2009 partial update for the 2010 edition, and references
therein.

\bibitem{baryogenesis}
A.D.\ Sakharov, Sov. Phys. JETP Lett. {\bf 5}, 24 (1967); 
P.\ Huet and E.\ Sather, Phys. Rev. D {\bf 51}, 379 (1995).

\bibitem{D0_pubs}
V.M.\ Abazov \etal\ (D\O\ Collaboration),
Phys. Rev. D, in press [arXiv:1005.2757]; 
V.M.\ Abazov \etal\ (D\O\ Collaboration),
Phys. Rev. Lett., in press [arXiv:1007.0395].

\bibitem{equala}
Y.\ Grossman, Y.\ Nir, and G.\ Raz,
Phys.\ Rev.\ Lett.\ {\bf 97}, 151801 (2006).

\bibitem{SMpred}
A.\ Lenz and U.\ Nierste,
J. High Energy Phys. {\bf 0706}, 072 (2007).

\bibitem{d0jpsiphi}
V.M.\ Abazov \etal\ (D\O\ Collaboration),
Phys. Rev. Lett. {\bf 101}, 241801 (2008).

\bibitem{kp}
V.A.\ Kosteleck\'y and R.\ Potting,
Phys.\ Rev.\ D {\bf 51}, 3923 (1995).

\bibitem{ksp}
V.A.\ Kosteleck\'y and S.\ Samuel,
Phys.\ Rev.\ D {\bf 39}, 683 (1989);
V.A.\ Kosteleck\'y and R.\ Potting,
Nucl.\ Phys.\ B {\bf 359}, 545 (1991).

\bibitem{d0cdf}
D\O\ Collaboration,
V.M.\ Abazov \etal,
Phys.\ Rev.\ Lett.\ {\bf 97}, 021802 (2006);
CDF Collaboration,
A.\ Abulencia \etal,
Phys.\ Rev.\ Lett.\  {\bf 97}, 242003 (2006).

\bibitem{ck}
D.\ Colladay and V.A.\ Kosteleck\'y,
Phys.\ Rev.\ D {\bf 55}, 6760 (1997);
Phys.\ Rev.\ D {\bf 58} 116002 (1998);
V.A.\ Kosteleck\'y,
Phys.\ Rev.\ D {\bf 69}, 105009 (2004).

\bibitem{ak3}
V.A.\ Kosteleck\'y,
Phys.\ Rev.\ D {\bf 64}, 076001 (2001).

\bibitem{owg}
O.W.\ Greenberg,
Phys.\ Rev.\ Lett.\ {\bf 89}, 231602 (2002).

\bibitem{cpt_preprint}
V.A.\ Kosteleck\'y and R.\ Van Kooten, arXiv:1007.5312.



\end{thebibliography}
\end{document}